\documentclass[letterpaper]{jpconf}
\usepackage{graphicx}
\begin{document}
\title{Observation of W decay in 500 GeV $p+p$ collisions at RHIC}

\author{Kensuke Okada for the PHENIX collaboration}

\address{RIKEN BNL Research Center}

\begin{abstract}
W boson production is observed in $\sqrt{s}=500$ GeV proton proton 
collisions at RHIC-PHENIX experiment. 
The single longitudinal spin asymmetry 
$A_L(\overrightarrow{p}p\rightarrow W^+X)$ is 
measured via decay positrons in the mid rapidity region.
The asymmetry $-0.83\pm0.31\pm(11\%$ scale uncertainty) is  
consistent with calculations from various polarized parton 
distribution functions.
\end{abstract}

\section{Introduction}
 The components of the proton spin is one of key questions in the nuclear physics.
It has been proposed to use W boson production to decompose anti quark 
contributions \cite{Wprogram}.  In leading order process, 
positive (negative) W boson is connected with u and $\bar{\rm d}$ (d and $\bar{\rm u}$) quarks. 
Because of the parity violating process, it provides a unique way to access the 
 flavor dependence of polarized quarks in the proton. 
It is a complementary approach to semi inclusive deep inelastic interaction experiments, 
where the quark flavor is identified only via the fragmentation process. 
In 2009, RHIC provided the first polarized proton proton collisions at $\sqrt{s}=500$GeV. 
The proton polarization is achieved to be 39\% for both beams. 

In this article, we report the observation of W boson production 
and the measurement of 
single longitudinal spin asymmetry of 
$W^+$ production at RHIC-PHENIX.

\section{RHIC-PHENIX}
 The data were collected by the PHENIX detector \cite{PHENIXNIM}.   
The central arm detector covers a range of $|\eta|<0.35$ in pseudo-rapidity and 2 times 90$^\circ$ in azimuth.  
The primary detector for this measurement is an electromagnetic calorimeter (EMCal). Each calorimeter tower covers $\Delta \eta \times \Delta \phi \sim 0.01 \times 0.01$. A tracking system consist of a drift chamber 
and a pad chamber is used to identify the charge sign. 
 Events were collected with the EMCal trigger, which is fully efficient 
 at 12 GeV of the transverse energy. The calibration of EMCal energy 
 scale was done with two photons' invariant mass for $\pi^0$ and $\eta$ 
 particles. For the tracking system, the detector position was
calibrated with zero magnetic field data. The angular resolution 
 of the tracking system was also checked with zero magnetic field data. 
 Compared with the bending angle from the magnetic filed, there is
 2.1 $\sigma$ charge separation capability for 40 GeV/$c$ particles.
A coincidence of beam-beam counters (BBC) positioned at pseudo-rapidities 
 $3.1<|\eta|<3.9$ was used as the luminosity monitor.
The conversion factor from the count rate to the luminosity
was obtained via the van der Meer scan technique 
\cite{vernierscan}, which measures the transverse profile of the beam 
overlap.
A correction was necessary for the event overlap in a single crossing. 
An integrated luminosity of 8.6 $\rm{pb}^{-1}$ after a vertex cut of 
$\pm 30$ cm is used in this analysis.

\section{W boson signal} 
\begin{figure}[h]
\includegraphics[width=18pc]{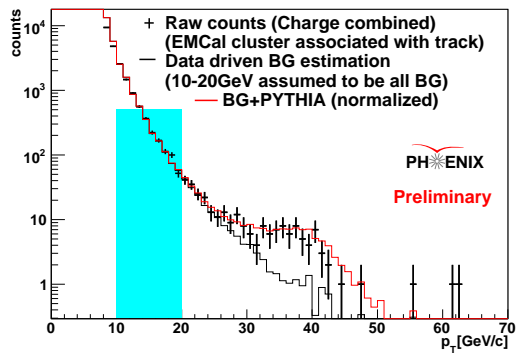}\hspace{2pc}%
\begin{minipage}[b]{18pc}\caption{\label{label} Transverse momentum spectra of EMCal clusters associated with a track in the PHENIX central arm detector. It is overlaid with curves of background and signal. The region of 10-20GeV is used for the normalization.}
\label{fig:spectra_pm}
\end{minipage}
\end{figure}

 In this analysis, W boson is tagged by its decay electron. 
 Electron candidates were selected from clusters in the EMCal within 
 $\pm 10$ mrad of the transverse position projected by a 
charged track found in the central drift chamber.
It applied loose cuts on the time measured in the EMCal and track momentum not to 
be much lower than the deposit energy in the EMCal.
Figure \ref{fig:spectra_pm} shows the transverse momentum spectra. 
The transverse momentum ($p_T$) is calculated from the energy deposit 
in the EMCal.
The histogram is overlaid with curves of background (QCD events) 
and signal (weak boson decays). 
Because decay electrons from Z bosons have 
similar spectra to the ones from W bosons, they can not be separated 
in the current PHENIX detector acceptance.
The main contributions from the QCD background are charged hadron 
clusters, photons from hadrons' decay 
converted to electrons before the tracking system, and some are from 
track mis-association in the same jet event. 
These backgrounds are estimated from all EMCal cluster distribution
for the photon contribution and the NLO pQCD distribution 
folded by the EMCal response for the charged hadron contribution. 
First the EMCal distribution was multiplied by the probability 
of track association. Second the charged hadron distribution was scaled, 
so that the range from 10 to 20 GeV/$c$ is explained by the sum of them.
%These backgrounds
%are estimated from all EMCal cluster distribution scaled 
%by the probability of track association and the rest is filled 
%by the shape 
%of charged pion distribution calculaed by the NLO pQCD folded by the 
%EMCal response, so that 
%the range from 10 to 20 GeV/$c$ is explained by the QCD enents. 
The signal shape 
is the decay electrons from W and Z bosons taken from PYTHIA MC 
smeared by the EMCal resolution.
The dominant systematic uncertainty in the background estimation is 
from the conversion probability. It has to 
estimate not only for a single photon conversion, but also for a probability of two photons, 
because two high energy clusters from $\pi^0$ decay merge in the EMCal.

A significant excess corresponding to the W boson Jacobian peak is observed in the spectra.

\section{Single longitudinal spin asymmetry}
\begin{figure}[h]
\begin{minipage}{18pc}
\includegraphics[width=18pc]{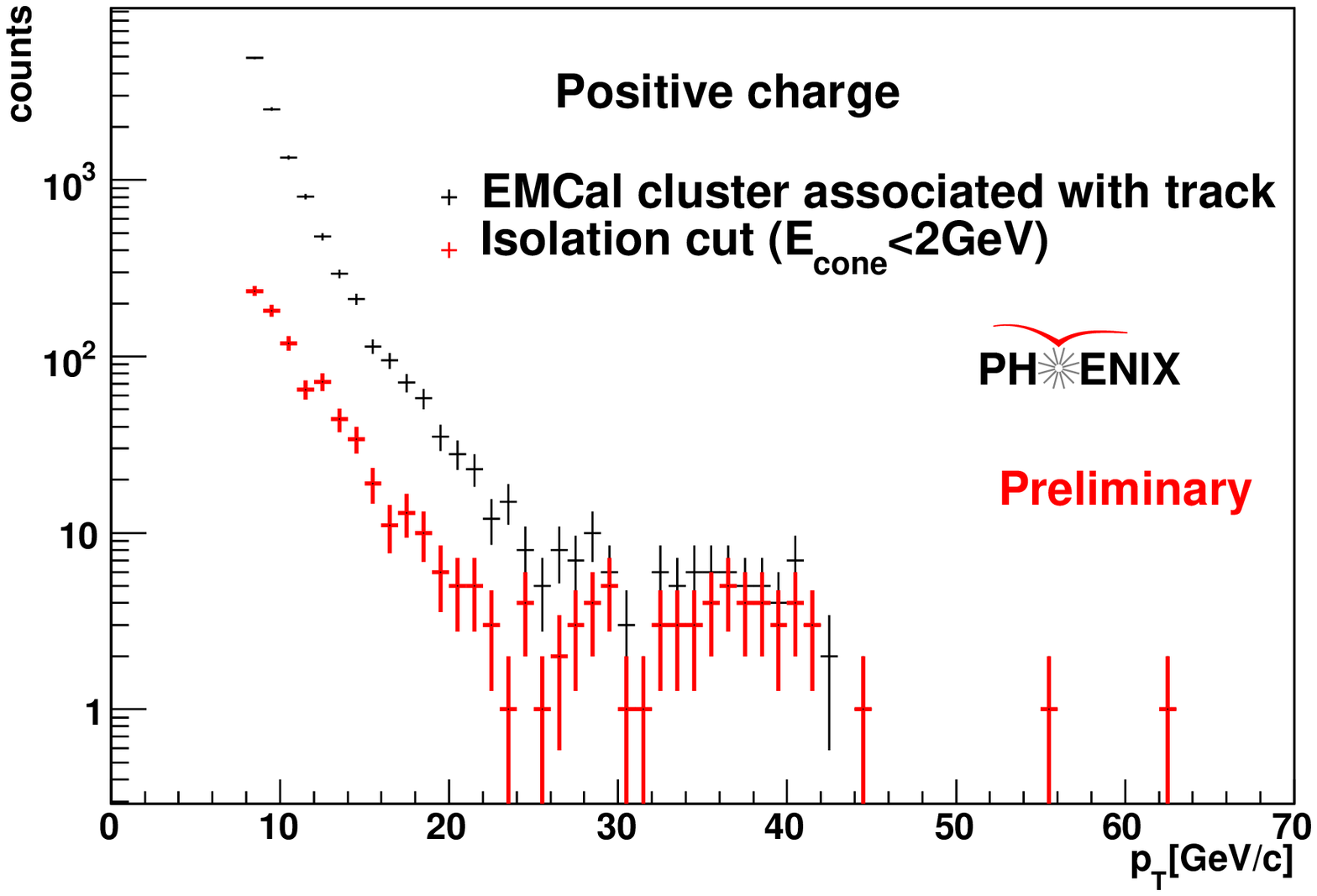}
\caption{\label{label} $p_T$ spectra of positive particles with and without the isolation cut.}
\label{fig:wiso_pos}
\end{minipage}\hspace{2pc}%
\begin{minipage}{18pc}
\includegraphics[width=18pc]{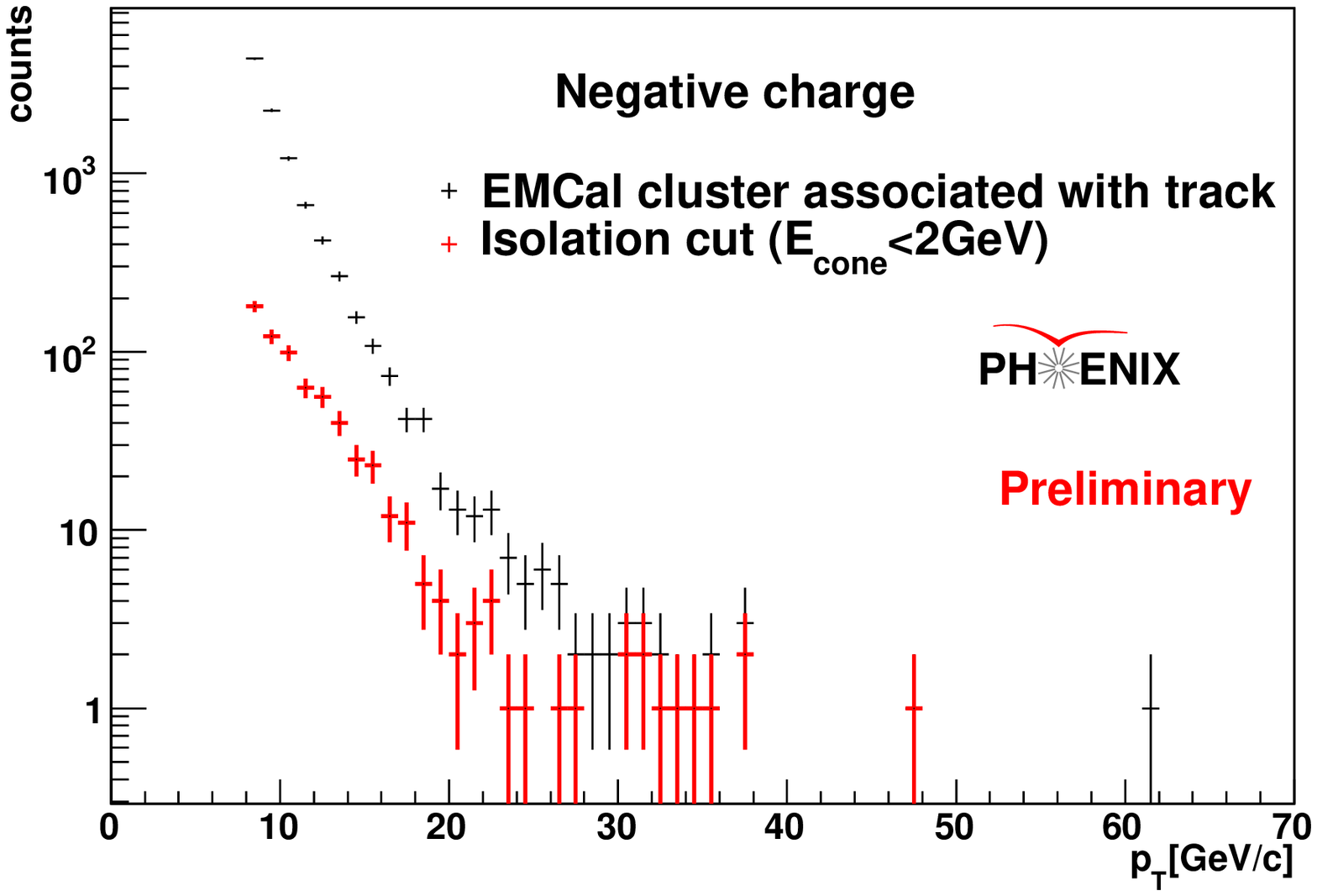}
\caption{\label{label} $p_T$ spectra of negative particles with and without the isolation cut.}
\label{fig:wiso_neg}
\end{minipage} 
\end{figure}

\begin{figure}[h]
\includegraphics[width=18pc]{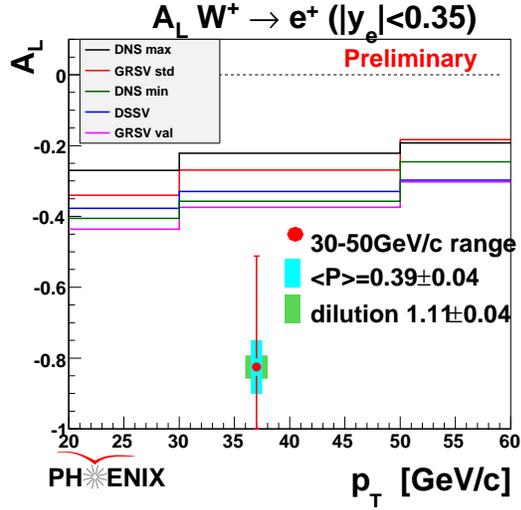}\hspace{2pc}%
\begin{minipage}[b]{18pc}\caption{\label{label} Longitudinal single spin asymmetry of $W^+ \rightarrow e^+$ in the mid rapidity region. It is compared with calculations from various polarized parton distribution functions.
(GRSV standard, GRSV valence \cite{GRSV}, DSSV \cite{DSSV}, and DNS \cite{DNS} using a maximal and minimal sea polarization scenario.)}
\label{fig:asym_plus}
\end{minipage}
\end{figure}

 For the spin asymmetry measurement, an isolation cut was applied 
 to increase the signal to noise ratio. The cut is to require 
less than 2 GeV of energy deposit around the electron candidate. 
This cut should not depend on the spin state.
 Figure \ref{fig:wiso_pos} and Fig. \ref{fig:wiso_neg} show the 
 spectra of inclusive and the one with 
 the isolation cut for positive and negative particles. 
 It is seen the background component from the QCD events 
 ($p_T<25{\rm GeV}/c$) is suppressed by a factor of about 4, 
 and the signal from W boson ($p_T>30{\rm GeV}/c$) is mostly remained. 
This is another evidence for the W boson signal is in the sample.
The single longitudinal spin asymmetry is defined as 
$A_L^W=\frac{1}{P}\cdot \frac{N^+(W)-N^-(W)}{N^+(W)+N^-(W)}$, 
where $P$ is the beam polarization, and $N^\pm$ is the number of 
signals normalized by the integrated luminosity
in positive and negative helicity beam.
For the luminosity measurement, 
number of BBC coincidence was used. 
Since there are two beams polarized at RHIC, 
the same sample can be used twice. Therefore the statistical uncertainty 
follows 
$\delta A_L=\frac{1}{P}\cdot\frac{1}{\sqrt{2(N^++N^-)}}$. 
In the actual calculation, the sample was divided into 4 spin states 
(2 beams $\times$ 2 spin states), then a 
simultaneous fit was applied to get raw asymmetries ($\equiv P\cdot A_L$).
Table \ref{table:rawasym} shows the raw asymmetry of positive particles. It also shows the background 
region for a sanity check expecting the asymmetry to be 0.
 For the physics asymmetry of W bosons, it has to be corrected for 
 the contribution of Z boson 
and QCD background in the sample. Those work as a dilution factor. 
The QCD background contribution is estimated to 
be $1\pm1$ event from the extrapolation of the lower part of 
the spectra.  
The ratio of Z boson to W boson is taken from PYTHIA MC ((W+Z)/W=1.08). 
Figure \ref{fig:asym_plus} shows the physics asymmetry of $W^+$ production after the correction.
The systematic uncertainty is from the absolute polarization measurement ($\delta P/P$=9.2\%) and 
the estimation of dilution effect ($\sim$4\%).  
The asymmetry is consistent with predictions of various polarized 
PDF parametrization within 
the uncertainty and it is 2.7$\sigma$ away from 0. 

\begin{table}
\centering
\caption{\label{book}Single longitudinal spin asymmetry for positive particles}
\begin{tabular}{@{}l*{15}{l}}
\br
$p_T$ Range [GeV/$c$]&Raw asymmetry \\
\mr
12-20 (Background) & $0.035\pm0.047$ \\
30-50 (Signal) & $-0.29\pm0.11$ \\
\br
\end{tabular}
\label{table:rawasym}
\end{table}

\section{Summary}
 The W boson production is observed through its decay electrons at 
RHIC-PHENIX in the data 
collected in 2009 ($\int{L}=8.6$/pb, P=39\%). The single longitudinal
spin asymmetry of $W^+$ boson is measured for the first time and 
it is consistent with various predictions within the uncertainty.

\section{Outlook}
 The RHIC spin program is planning to accumulate $\sqrt{s}=500$ GeV 
 collision data for the next few years. There is also a plan to 
 measure W boson to muon channel in the forward muon detectors. 
 It has better 
 sensitivity to the polarized $\bar{\rm u}$ distribution than the central 
 arm measurement because of its decay kinematics. 
 A detector upgrade to improve the trigger capability 
is the major challenge in the next data taking.

 The measurement of beam polarization will be an 
important systematics for non-zero helicity asymmetries.
The systematic uncertainty has been already achieved at the level of 
$\sim5\%$ in the absolute polarization measurement in the past 
200GeV data periods. It is expected to be the same level in the next 
500GeV data period.


\section*{References}
\begin{thebibliography}{9}
\bibitem{Wprogram} G. Bunce, N. Saito, J. Soffer, and W. Vogelsang, Ann. Rev. Nucl. Part. Sci. {\bf 50}, 525 (2000), hep-ph/0007218.
\bibitem{PHENIXNIM} K.~Adcox et al.: Nucl. Inst. Meth. A{\bf 499}, 469 (2003).
\bibitem{vernierscan} K. A. Drees and Z. Xu, "Proceedings of the PAC2001 Conference," 3120 (2001).
\bibitem{GRSV}M. Gluck, E. Reya, M. Stratmann, and W. Vogelsang, Phys. Rev. {\bf D63}, 094005 (2001), hep-ph/0011215.
\bibitem{DSSV} D. de Florian, R. Sassot, M. Stratmann, and W. Vogelsang,  Phys. Rev. Lett. {\bf 101},072001 (2008), hep-ph/0804.0422.
\bibitem{DNS} D. de Florian, G. A. Navarro, and R. Sassot, Phys. Rev. {\bf D71}, 094018 (2005), hep-ph/0504155.
\end{thebibliography}
\end{document}